\documentclass{article}

\begin{document}

\title{A master equation approach for the interaction of an atom with a dielectric semi-infinite medium}         
\author{T.N.C. Mendes\footnote{tarciro@if.ufrj.br} $\;$ and C. Farina\footnote{farina@if.ufrj.br}}        
\maketitle

\begin{abstract}
We use the master equation approach to calculate the energy level shifts of an atom in the presence of a general dielectric semi-infinite medium characterized by a dielectric constant $\epsilon(\omega)$. Particularly, we analyze the case of a non-dispersive medium for which we obtain a general expression for the interaction as well as the asymptotic behaviors for $k_0 z \ll 1$ (non-retarded regime) and  $k_0 z \gg 1$ (retarded regime), where $\omega_0 = k_0 c$ is the main transition frequency of the atom. The limiting cases $\epsilon \simeq 1$ and $\epsilon \gg 1$ are discussed for both retarded and non-retarded limits. For the retarded limit, we compute the non-additivity contribution of van der Waals forces.
\end{abstract}

\section{Introduction}       

Experiments performed by Verwey and Overbeek \cite{Overbeek} with colloidal suspensions during the years 1940-45
suggested that London's \cite{London} interaction between two polarizable atoms (that falls as $1/r^6$) was not correct for large distances. In order to get an agreement between theory and experimental data the van der Waals force should fall faster
than $1/r^6$ for large distances. The authors conjectured that this change in the behavior of the force was due to the retardation effects of the electromagnetic interaction or, in other words, due to the finiteness of the velocity of light.

Motivated by the discrepancies pointed out by Verwey and Overbeek, in 1948 Casimir and Polder \cite{CasiPolder} considered for the first time the influence of retardation effects on the van der Waals forces between two atoms and between an atom and a perfectly conducting wall. The non-retarded dispersive van der Walls force between a polarizable atom and a perfectly conducting wall can be explained with the aid of the image method \cite{CCT_QM}, which shows that the interaction potential varies as $1/z^3$, where $z$ is the distance between the atom and the wall. When retardation effects are taken into account, Casimir and Polder \cite{CasiPolder} showed using perturbative QED that the interaction potential becomes proportional to $1/z^4$. Since Casimir and Polder's paper, forces between atoms or molecules and any kind of walls are usually called \textit{Casimir-Polder forces}.

In 1956, Lifshitz and collaborators \cite{Lifshitz} developed a general theory of van der Waals forces. They derived a powerful expression for the force at finite temperature between two semi-infinite dispersive media characterized by well defined dielectric constants and separated by a slab of any other dispersive medium. Several results have been
predicted with the aid of this formula, as for example, the variation of the thickness of thin superfluid helium films in a remarkable agreement with the experiments \cite{SabAnd}. The Casimir-Polder force can also be obtained as a limiting case of Lifshitz formula when one of the media is sufficiently dilute such that the force between the slabs may be obtained by direct integration of single atom-wall interactions \cite{DLP}.

Since then, a lot of work has been done on van der Waals and Casimir-Polder interactions. Good reviews have been written on dispersive van der Waals interaction \cite{Milonni,Langbein,Margenau} and many  elaborated papers concerning level-shifts near surfaces have appeared, as for instance, \cite{Meschede,HindsSandoghdar,Jhe,NhaJhe}, to mention just a few. It is worth mentioning that Casimir-Polder forces have been observed experimentally \cite{HindsetAl93}. Recently, the influence of real conditions on the Casimir-Polder interaction has been considered \cite{Mostepanenko2004}. Further, higher multipole corrections \cite{Salam}, roughness  and corrugation of surfaces \cite{PAm,Emig}, the influence of the Casimir-Polder interaction on Bose-Einstein condensates \cite{Vuletic,Antezza} and possible applications to nanotubes \cite{Nanotubo} are some of the many branches of great activity on this subject nowadays.

Here we shall compute the van der Waals interaction between an atom and a dielectric semi-infinite medium characterized by a well defined dielectric constant $\epsilon(\omega)$. We shall employ the master equation approach \cite{DDC1984} recently applied by the authors \cite{TarciroJPA2006} to this kind of calculation. For the particular case of a non-dispersive medium ($\epsilon\left(\omega\right)=\epsilon=cte$ for any frequency), we will also compute in a closed form the variation of the interaction potential with $\epsilon$ in London-van der Waals and Casimir-Polder limits. A measure of the non-additivity of van der Waals forces will be given in the large distance regime (Casimir-Polder limit).
%
%
\section{General Level Shifts}

In a previous paper \cite{TarciroJPA2006} we analysed the interaction of an atom with a perfectly conducting wall starting from the general expressions for the energy level shifts of a small system interacting with a large one  considered as a reservoir. These expressions have been derived from so called {\it master equation} \cite{AtPhInt}. Let us briefly sumarize the formalism to be used.

Consider a system $\mathcal S + \mathcal R$, where by assumption $\mathcal S$ is a small system and $\mathcal R$ a reservoir. Starting from an interacting Hamiltonian of the form
\begin{equation}
\label{Vtdep}
V(t)=-\sum_j S_j(t)R_j(t)
\end{equation}
, where $S_j(t)$ and $R_j(t)$ are $\mathcal S$ and $\mathcal R$ observables, the master equation comes from the perturbative expansion of the density matrix of the total system $\mathcal S + \mathcal R$. Working with a second order perturbation theory, one can take the reduced trace over the Hilbert space of $\mathcal R$ and do the Markovian approximation. This approximation is due to the existence of two very different time scales, namely: the correlation time of the fluctuations of the $\mathcal R$ variables and the characteristic time evolution of $\mathcal S$, which is too large compared to the former. The final equation describes the evolution of the density matrix of $\mathcal S$, made up of a free term plus a linear term that accounts for the coupling with $\mathcal R$ \cite{AtPhInt}. The master equation is given by:
\begin{eqnarray}
\label{eqmestra}
\frac{d\rho_{ab}^S\left( t\right) }{dt} &=& -\; i\omega_{ab}\rho_{ab}^S\left( t\right)+\sum_{c,d}{\mathcal J}_{abcd}\rho_{cd}^S\left( t\right)
\\
%
\label{txmestra}
{\mathcal J}_{abcd}&=&-\;\frac{1}{\hbar^2}\sum_{j,k} \int_0^\infty d\tau
\Biggl\{g_{jk}\left( \tau\right) \left[ \delta_{bd} \sum_n S_{an}^j S_{nc}^k e^{-i\omega_{nc}\tau} - S_{ac}^k S_{db}^j e^{-i\omega_{ac}\tau}\right]
%
\nonumber \\ 
&+&\ g_{kj}\left( -\tau\right) \left[ \delta_{ac} \sum_n S_{dn}^k S_{nb}^j e^{-i\omega_{dn}\tau} - S_{ac}^j S_{db}^k e^{-i\omega_{db}\tau}\right]\Biggr\} 
\end{eqnarray}
where $g_{jk}\left(\tau\right) = \textrm{Tr}_R\left[\rho_R R_j\left(\tau\right) R_k\left( 0\right)\right]$,
$\rho_{R\left(S\right)}=\textrm{Tr}_{S\left(R\right)}\left(\rho_S\otimes\rho_R\right)$
is the reduced density matrix of $\mathcal R \left(\mathcal S \right)$, $\rho_{ab}^S = \langle a \vert \rho_S \vert b \rangle$, $S_{ab}^j = \langle a \vert S_j \vert b \rangle$, $\hbar\omega_{ab}=E_a-E_b$ and $\vert a\rangle$ represents an energy eigenstate of $\mathcal S$ with eigenvalue $E_a$. In order to obtain the level shifts of $\mathcal S$ it is necessary to consider only the non-diagonal terms which are coupled with themselves or, in other words, the terms in which the indexes $c = a$ and $d = b$. For simplicity, one assume that levels $\vert a \rangle$ and $\vert b \rangle$ are non-degenerate. Hence, from eq.(\ref{txmestra}), one may write:
\begin{eqnarray}
\label{levelshift}
{\mathcal J}_{abab}=-\Gamma_{ab} - i\Delta_{ab} \ ; \ \ 
\Delta_{a}=\frac{1}{\hbar}\sum_{\mu} p_{\mu} \sum_{\nu} \sum_n {\mathcal P}\frac{\vert\langle\nu,n\vert V\vert \mu, a\rangle\vert^2}{E_\mu+E_a-E_\nu-E_n}
\end{eqnarray}
where $\Delta_{ab} = \Delta_{a} - \Delta_{b}$, $\Gamma_{ab}$ is a damping, $\mathcal P$ is the Cauchy principal value, $p_\mu$ is the statistical weight of the state $\vert \mu \rangle$ which is an eigenstate of $\mathcal R$ with energy eigenvalue $E_\mu$ and $V$ is defined by eq.(\ref{Vtdep}). Using the definitions of the linear susceptibility and symmetric correlation function for both small system and reservoir, the level shifts can be splited into two terms: the first one is due to polarization of the system induced by the fluctuations of reservoir $(fr)$ and the other is the contribution of the reservoir reaction $(rr)$ over the system due to the fluctuations of the system induced by the reservoir. The expressions are:
\begin{eqnarray}
\label{fr}
\hbar \Delta_a^{fr}\!\!\!\!&=&\!\!\!\!-\frac{1}{2}\sum_{j,k}\int_{-\infty}^{\infty}\frac{d\omega}{2\pi}\chi_{jk}^{\prime S,a}\left( \omega\right) C_{jk}^{R}\left( \omega\right) 
\\
%
\label{rr}
\hbar \Delta_a^{rr}\!\!\!\!&=&\!\!\!\!-\frac{1}{2}\sum_{j,k}\int_{-\infty}^{\infty}\frac{d\omega}{2\pi}\chi_{jk}^{\prime R}\left( \omega\right) C_{jk}^{S,a}\left( \omega\right)
\\
%
\label{CSa}
C_{jk}^{S,a}\left( \omega\right) \!\!\!\!&=&\!\!\!\! \pi \sum_{n} S_{a n}^{j} S_{n a}^{k}\left[ \delta\left( \omega+\omega_{an}\right) +\delta\left( \omega-\omega_{an}\right) \right]
\\
%
\label{XSa}
\chi_{jk}^{\prime S,a}\left( \omega\right)\!\!\!\! &=&\!\!\!\! -\frac{1}{\hbar}\sum_{n} S_{a n}^{j} S_{n a}^{k}\left[ {\mathcal P}\frac{1}{\omega_{a n}+\omega}+{\mathcal P}\frac{1}{\omega_{a n}-\omega}\right] 
\\
%
\label{CR}
C_{jk}^{R}\left( \omega\right)\!\!\! &=&\!\!\! \pi \sum_{\mu}p_{\mu}\sum_{\nu}R_{\mu\nu}^{j} R_{\nu\mu}^{k}\left[ \delta\left( \omega+\omega_{\mu\nu}\right) +\delta\left( \omega-\omega_{\mu\nu}\right) \right]
\\
%
\label{XR}
\chi_{jk}^{\prime R}\left( \omega\right)\!\!\! &=&\!\!\! -\frac{1}{\hbar}\sum_{\mu}p_{\mu}\sum_{\nu}R_{\mu\nu}^{j} R_{\nu\mu}^{k}\left[ {\mathcal P}\frac{1}{\omega_{\mu\nu}+\omega}+{\mathcal P}\frac{1}{\omega_{\mu\nu}-\omega}\right] 
\end{eqnarray}
where $\delta E_a = \hbar\Delta_a=\hbar\Delta_a^{fr}+\hbar\Delta_a^{rr}$ is the energy level shift of $\mathcal S$ in the energy eigenstate $\vert a \rangle$ with eigenvalue $E_a$, $\chi_{jk}^{\prime R}\left( \omega\right)$ is the real part of the susceptibility of $\mathcal R$ and $C_{jk}^{R}\left( \omega\right)$ is its symmetric correlation function, $C_{jk}^{S,a}\left( \omega\right)$ and $\chi_{jk}^{\prime S,a}\left( \omega\right)$ are the equivalents of $C_{jk}^{R}\left( \omega\right)$ and $\chi_{jk}^{\prime R}\left( \omega\right)$ for $\mathcal S$ in the state $\vert a \rangle$ and $R_{\mu\nu}^{j} = \langle \mu \vert R_j \vert \nu \rangle$, where $\vert \mu \rangle$ is the energy eigenstate of $\mathcal R$ with an eigenvalue $E_{\mu}$.

Let us consider the electromagnectic field as the reservoir and the atom as the small system. This is allways possible when the correlation in time of the field observables is very sharp compared to the correlation of the atom observables: this last correlation width is $\sim 1/\omega_0$ while the field correlation is $\sim 1/\omega_c$, where $\omega_0$ is the main frequency transition of the atom and $\omega_c$ is an arbitrarily large cut-off frequency.

In the dipole approximation the interacting Hamiltonian may be written as:
\begin{equation}
V({\bf x},t) = -{\bf d}\left( t\right) \cdot {\bf E}\left( {\bf x}, t\right)
\end{equation}
where ${\bf d}\left( t\right)$ is the dipole moment (the system observable) of the atom induced by the electric field (the reservoir obervable). Since this Hamiltonian is bilinear in the atom and field operators, one may use expressions (\ref{fr}) and (\ref{rr}) to calculate the interaction.

Consider a field mode as %
\begin{equation}
\label{Efield}
{\bf E}_{{\bf k}\lambda}\left({\bf x},t\right) = {\bf F}_{{\bf k}\lambda}\left({\bf x},t\right)a_{{\bf k}\lambda}^{\dag}+h.c.
\end{equation}
(this is always possible in a linear medium), where ${\bf F}_{{\bf k}\lambda}\left({\bf x},t\right)$ is a function of time and position that takes into account all sourced contributions and boundary conditions imposed to the field, and $a_{{\bf k}\lambda}^{\dag}$ is the creation operator of a photon with wave-vector ${\bf k}$ and polarization $\lambda$ which satisfies the commutation relations $\left[a_{{\bf k}\lambda},a_{{\bf k}^{\prime}\lambda^{\prime}}^{\dag}\right]=\delta_{{\bf k}{\bf k}^{\prime}}\delta_{\lambda\lambda^{\prime}}$. Consider also the atom as an isotropic two-level system with transition frequency $\omega_0=k_0 c=(E_e-E_g)/\hbar$ where $E_e$ and $E_g$ are the energy of its unperturbed excited $\vert e\rangle$ and fundamental $\vert g\rangle$ states respectively. This implies that all non-diagonal terms of its susceptibility and second order symmetric correlation function given by (\ref{XSa}) and (\ref{CSa}), respectively, are identically zero or, in other w!
 ords: $\chi_{jk}^{S,a}\left(\omega\right)=\chi^{a}\left(\omega\right)\delta_{jk}$ and $C_{jk}^{S,a}\left(\omega\right)=C^{a}\left(\omega\right)\delta_{jk}$. Then, if the atom is in its ground-state, equations (\ref{CSa}-\ref{XR}) and (\ref{Efield}) will lead to:
\begin{eqnarray}
\label{XS0}
\chi^{\prime a}\left( \omega\right)&=&\frac{\alpha_0 k_0 c}{2}{\mathcal P} \left[\frac{1}{k_0 c+\omega} + \frac{1}{k_0 c-\omega}\right]
\\
%
\label{CS0}
C^a\left( \omega\right)&=& \frac{\pi}{2}\hbar  k_0 c \alpha_0 \Bigl[ \delta\left( k_0 c  + \omega\right) + \delta\left( k_0 c - \omega\right)\Bigr]
\\
%
\label{XR0}
\chi^{\prime R}_{{\bf k}\lambda}\left(\omega\right)&=& {1\over \hbar} \vert {\bf F}_{{\bf k}\lambda}\left({\bf x},t\right)\vert^2 \left[{\mathcal P} \frac{1}{kc + \omega} + {\mathcal P} \frac{1}{kc - \omega}\right]
\\
%
\label{CR0}
C^{R}_{{\bf k}\lambda}\left( \omega\right)&=&2 \pi \vert {\bf F}_{{\bf k}\lambda}\left({\bf x},t\right)\vert^2 \left( \langle n_{{\bf k}\lambda}\rangle+{1\over 2}\right)  \Bigl[ \delta\left( k c  + \omega\right) + \delta\left( k c - \omega\right)\Bigr],
\end{eqnarray}
where $\alpha_0=2\vert {\mathbf d}_{eg}\vert^2/3\hbar\omega_0$ is the static polarizability, $k=\vert {\bf k}\vert$ and $\langle n_{{\bf k}\lambda}\rangle$ is the statistical average number of photons in a given mode. Combining equations (\ref{XS0}-\ref{CR0}) with (\ref{fr}) and (\ref{rr}), one can write the level shift $\delta E=\hbar \delta \omega_0=\delta E^{fr}+\delta E^{rr}$ as:
\begin{eqnarray}
\label{Level_rr}
\delta E^{rr}&=&-{1\over 2}\sum_{{\bf k}\lambda}\alpha_{-}(k)\vert {\bf F}_{{\bf k}\lambda}\left({\bf x},t\right)\vert^2
\\
%
\label{Level_fr}
\delta E^{fr}&=&-\sum_{{\bf k}\lambda}\alpha_{+}(k)\vert {\bf F}_{{\bf k}\lambda}\left({\bf x},t\right)\vert^2\left(\langle n_{{\bf k}\lambda}\rangle+{1\over 2}\right)
\\
%
\label{alpha}
\alpha_{\mp}\left( k\right)&=&\frac{\alpha_0 k_0}{2}\left(  {\mathcal P}\frac{1}{k + k_0} \pm {\mathcal P}\frac{1}{k - k_0}\right) .
\end{eqnarray}

In equation (\ref{Level_fr}) $\alpha_{+}\left( k\right)$ is the well known approximation for the polarizability of a two-level system with a negligible natural line-width $\Gamma_n$, namely $\omega_0\gg \Gamma_n$. It is formed by a resonant part, $\mathcal P \left[(k-k_0)^{-1}\right]$, and a non-resonant one, $\mathcal P \left[(k+k_0)^{-1}\right]$. Then, the dipole moment ${\bf d}_{\mathcal S}$ of system $\mathcal S$ induced by the field is given by ${\bf d}_{\mathcal S}\left({\bf x},t\right)=\sum_{{\bf k}\lambda}\alpha_{+}\left( k\right){\bf E}_{{\bf k}\lambda}\left({\bf x},t\right)$.

By the same argument, $\alpha_{-}\left( k\right)$ may be interpreted as a field polarizability related to the reaction on itself of its influence on the system $\mathcal S$. Equivalently, one may talk about the
\lq\lq dipole moment{\rq\rq} ${\bf d}_{\mathcal R}$ of the field, that can be written as ${\bf d}_{\mathcal R}\left({\bf x},t\right)=\sum_{{\bf k}\lambda}\alpha_{-}\left( k\right){\bf E}_{{\bf k}\lambda}\left({\bf x},t\right)$.

For the vacuum state of the field we have $\langle n_{{\bf k}\lambda}\rangle=0$ and the total energy level-shift may be written as:
\begin{equation}
\label{tot_level}
\delta E = -\frac{\alpha_0 k_0}{2}\mathcal P\sum_{{\bf k}\lambda}\frac{\vert {\bf F}_{{\bf k}\lambda}\left({\bf x},t\right)\vert^2}{k+k_0}
\end{equation}
where only the non-resonant term contributes. In the following section we will apply the previous equation to find a general expression for the interaction between an atom and a semi-infinity dispersive dielectric medium for all distance regimes and all possible values of the dielectric constant.
%
%
\section{Force between the atom and the dielectric medium}

Let us consider a semi-infinite dielectric medium with a dielectric constant $\epsilon\left(k\right)$ defined in the region $z \leq 0$. Let us also consider that there exists an atom at a position $z>0$. For this simple system, it is possible to separate explicitly the modes of the electromagnetic field \cite{Carniglia}. Hence, for the electric field one may write  \cite{Milonni}:
\begin{eqnarray}
{\bf E}_{{\bf k}\lambda_{1(2)}}\left({\bf x},t\right) &=& -i\left({2\pi\hbar c k \over V}\right)^{1/2}{\bf A}_{{\bf k}\lambda_{1(2)}}\left({\bf x}\right) e^{i k c t} a_{{\bf k}\lambda_{1(2)}}^{\dag}(0)+h.c.
\\
%
{\bf A}_{{\bf k}\lambda_1}\left({\bf x}\right) &=& {1\over \sqrt{2}}e^{-i{\bf k}\cdot {\bf x}}{\bf e}_{{\bf k}\lambda}\left[1+\left(k_3-k_3^{\prime} \over k_3+k_3^{\prime}\right)e^{2ik_3 z}\right]
\\
%
{\bf A}_{{\bf k}\lambda_2}\left({\bf x}\right) &=& {1\over k\sqrt{2}}e^{-i{\bf k}\cdot {\bf x}}{\bf e}_{{\bf k}\lambda}\times\left[{\bf k}+ {\bf k^{\prime}}\left({\epsilon\left(k\right)k_3-k_3^{\prime} \over \epsilon\left(k\right)k_3+k_3^{\prime}}\right)e^{2ik_3 z}\right]
\\
%
k_3 &=& \sqrt{k^2-k_{\bot}^2}\;\; ; \;\; k_3^{\prime} = \sqrt{\epsilon\left(k\right)k^2-k_{\bot}^2} \;\; ; \;\; 
\\
%
{\bf k} &=& {\bf k}_{\bot} + k_3 \hat z \;\; ; \;\; {\bf k^{\prime}} = {\bf k}_{\bot} - k_3 \hat z \;\;\; \textrm{and} \;\;\; {\bf k}_{\bot}=k_1 \hat x+k_2 \hat y
\end{eqnarray}
As a consequence, the sum in polarizations of the modulus square of the amplitude of the field is given by:
\begin{eqnarray}
\sum_{\lambda}\vert {\bf F}_{{\bf k}\lambda}\left({\bf x},t\right)\vert^2 &=& {2\pi \hbar k c \over V}\Bigg\lbrace 1+{1\over 2}\left({k_3-k^{\prime}_3\over k_3+k^{\prime}_3}\right)^2+{1\over 2}\left({\epsilon\left(k\right)k_3-k^{\prime}_3\over \epsilon\left(k\right)k_3+k^{\prime}_3}\right)^2
\nonumber \\
&+& \left[{k_3-k^{\prime}_3\over k_3+k^{\prime}_3}+\left(2{k_{\bot}^2\over k^2}-1\right)\left({\epsilon\left(k\right)k_3-k^{\prime}_3\over \epsilon\left(k\right)k_3+k^{\prime}_3}\right)\right]\cos\left(2k_3 z\right)\Bigg\rbrace
\end{eqnarray}
Using the previous equation on (\ref{tot_level}), remembering that $\sum_{{\bf k}}\rightarrow V/\left(2\pi\right)^3\int k^2dk\int d\Omega$ in spherical coordinates and keeping only the $z$-dependent part of $\delta E$, which is the only one that contributes to the interaction of the atom with the wall, one obtains after some mathematical manipulations:
\begin{eqnarray}
\label{Vepsk}
\delta E\left(z\right)&=&V\left(z\right)=-{\hbar c \over 4\pi}\alpha_0 k_0\int_0^{\infty}\! dk\;{k^3 \over k+k_0}\int_{-1}^1 dt f\left(t,\epsilon(k)\right)\cos\!\left(2kzt\right)
\\
\label{fepsk}
f\left(t,\epsilon(k)\right) &=& {\vert t\vert -\sqrt{\epsilon\left(k\right)-1+t^2} \over \vert t\vert +\sqrt{\epsilon\left(k\right)-1+t^2}} + \left(1-2t^2\right){\vert t\vert \epsilon\left(k\right)-\sqrt{\epsilon\left(k\right)-1+t^2} \over \vert t\vert \epsilon\left(k\right)+\sqrt{\epsilon\left(k\right)-1+t^2}}
\end{eqnarray}
where the principle part $\mathcal P$ was omited because there is no poles, since $k=\vert {\bf k}\vert$ and $k_0$ are allways positive and the integral in $k$ is over on the positive real axis. 

Equations (\ref{Vepsk}-\ref{fepsk}) can be considered one of the main results of the present paper. They give the interaction between an atom and a semi-infinite dielectric medium for all distance regimes and any frequency dependence of the dielectric constant. As an important particular case we shall consider in a moment a non-dispersive medium. It is
worth mentioning that Nha and Jhe \cite{NhaJhe} have also considered non-dispersive media. They calculated the level shifts of an atom between two parallel dielectric surfaces using linear response theory \cite{Kubo1966,McLT=0}.

Similar results involving dispersive media have been deduced in the literature directly from the interatomic potentials or from Lifshitz formula (see \cite{Milonni} and references in it). However, as far we know, in all of them it is necessary to consider at least one rarefied medium and, if necessary, add perturbatively the corrections  due to the non-additivity of the van der Waals forces, which arises from many-body interatomic potentials \cite{PowerThiru1985,PowerThiru1994} (for a simple discussion see \cite{Farina}).
Note that our result gives a very compact formula for the interaction between the atom and the semi-infinite dielectric medium which includes all many-body interactions.

In order to proceed, it is necessary to have the explicity form of $\epsilon\left(k\right)$. Let us analyze the particular case of a non-dispersive medium, where the dielectric constant may be considered independent of $k$ in a large range of frequencies or, in other words: $\epsilon\left(k\right)\simeq \epsilon\left(0\right) =\epsilon$. For this case, one can  make firstly the integration on $k$ in equation (\ref{Vepsk}), which leads to
\begin{eqnarray}
\label{VepsF3}
V\left(z,\epsilon\right)&=& -{\hbar c \over 32\pi}{\alpha_0 k_0\over z^3}\int_{-1}^1 dt f\left(t,\epsilon\right)\int_0^{\infty}\! dx\;{x^3 \cos\left(x t\right)\over x+x_0}
\nonumber \\
&=& {\hbar c \over 4\pi}\alpha_0 k_0^4\int_{-1}^1 dt f\left(t,\epsilon\right)\mathcal F^{\,\prime\prime\prime}\left(x_0 t\right)
\\
%
\mathcal F\left(x_0 t\right)&=&\int_0^{\infty}dx{\sin x t \over x+x_0}=\textrm{Ci}\left( x_0 t\right) \sin x_0 t - \textrm{si}\left( x_0 t\right)\cos x_0 t 
\\
%
\mathcal G\left(x_0 t\right)&=&\mathcal F^{\prime}\left(x_0 t\right)=-\int_0^{\infty}dx{\cos x t \over x+x_0}=\textrm{Ci}\left( x_0 t\right) \cos x_0 t + \textrm{si}\left( x_0 t\right)\sin x_0 t 
\\
%
\mathcal F^{(2n)}\left(x\right) &=& \mathcal G^{(2n-1)}\left(x\right)=\left(-1\right)^n \left[\mathcal F\left(x\right)-\sum_{j=0}^{n-1}{\left(-1\right)^j\left(2 j\right)! \over x^{2j+1}}\right]\;\;\;\;\; \textrm{with} \;\;\; n=1,2,3,...
\\
%
%
\mathcal F^{(2n+1)}\left(x\right) &=& \mathcal G^{(2n)}\left(x\right)=\left(-1\right)^n \left[\mathcal G\left(x\right)+\sum_{j=0}^{n-1}{\left(-1\right)^j\left(2 j+1\right)! \over x^{2j+2}}\right]\;\;\; \textrm{with} \;\;\; n=1,2,3,...
\\
%
\textrm{Ci}\left( x\right)&=& \gamma +\ln x + \int_0^x dt\; \frac{\cos t-1}{t}\;\;\;\;\;  \textrm{and}\;\;\;\;\; \textrm{si}\left( x\right) = -{\pi \over 2} +\int_0^x dt\; \frac{\sin t}{t}\; ,
%
%
\end{eqnarray}
where $\gamma$ is the Euler-Mascheroni constant and $x_0=2k_0z$. Integrating equation (\ref{VepsF3}) by parts, one  may write:
\begin{eqnarray}
\label{Vepsf3}
V\left(z,\epsilon\right) &=& -{\hbar c \over 16\pi}{\alpha_0 k_0 \over z^3} \bigg\lbrace \left[f\left(1,\epsilon\right)x_0^2-f^{\,\prime\prime}\left(1,\epsilon\right)\right]\mathcal F\left(x_0\right)+f^{\,\prime}\left(1,\epsilon\right)x_0\mathcal G\left(x_0\right)+ %
\nonumber \\
&-& x_0f\left(1,\epsilon\right)+ {1\over 2}\int_{-1}^1 dt f^{\,\prime\prime\prime}\left(t,\epsilon\right)\mathcal F\left(x_0 t\right)\bigg\rbrace
\end{eqnarray}
where $f^{\,\prime}$, $f^{\,\prime\prime}$, ...,  mean first, second, ..., partial derivatives of $f$ respect to $t$.
In order to check the self-consistency of our results, let us consider the particular cases $\epsilon=1$ and $\epsilon \rightarrow\infty$. By direct inspection of equation (\ref{fepsk}), we get
\begin{eqnarray}
\label{lim0}
\lim_{\epsilon\rightarrow 1}f(t,\epsilon)&=&0
\\
%
\label{limInfty}
\lim_{\epsilon\rightarrow \infty}f(t,\epsilon)&=&-2t^2
\end{eqnarray}
>From equations (\ref{lim0}) and (\ref{VepsF3}) we see that for $\epsilon =1$,  the  interaction potential is identically zero, as expected, since $\epsilon =1$ means that there is no medium at all.

For the case $\epsilon\rightarrow\infty$, the last term on the r.h.s. of equation (\ref{Vepsf3}) vanishes, since from (\ref{limInfty})
 $f^{\prime\prime\prime}\left(t,\epsilon\right)=0$. The remaining terms give a finite value so that the interaction
 potential may be written as:
\begin{equation}
\label{pcw}
V\left(z,\infty\right)=V_0\left(z\right)=\frac{\hbar c}{8 \pi} \frac{k_0 \alpha_0}{z^3} \left[\left( x_0^2 -2\right) {\mathcal F} \left( x_0\right) + 2 x_0 {\mathcal G}\left( x_0\right) - x_0\right]\, ,
\end{equation}
in perfect agreement with a previous result \cite{TarciroJPA2006}. Using the limiting expressions of $\mathcal F\left(x\right)$ and $\mathcal G\left(x\right)=\mathcal F^{\prime}\left(x\right)$ for small and large values of $x$:
\begin{eqnarray}
\label{FiS}
\mathcal F\left(x\right) &=& {\pi \over 2}-\left(1-\gamma \right)x + x\ln x + \mathcal O\left(x^2\right)\;\;\;\;\;\;\;\; (x \ll 1)
\\
%
\label{FiL}
\mathcal F\left(x\right) &\simeq& {1\over x}- {2\over x^3}+ {24\over x^5} - {720\over x^7}+... \;\;\;\;\;\;\;\;\;\;\;\;\;\;\;\;\;\;\; (x \gg 1)
\end{eqnarray}
one may write for the asymptotic behaviors of $V_0\left(z\right)$:
\begin{eqnarray}
\label{V0short}
V_0\left( z\right) &=& -\frac{\hbar \omega_0}{8} \frac{\alpha_0}{z^3} + {\mathcal O}\left( z^{-2}\right)\;\;\;\;\;\;\;\; \textrm{for} \;\;\; x_0 \ll 1
\\
%
\label{V0large}
V_0\left( z\right) &=& -\frac{3}{8\pi}\frac{\alpha_0\hbar c}{z^4} + {\mathcal O}\left( z^{-6}\right)\;\;\;\;\;\; \textrm{for} \;\;\; x_0 \gg 1
\end{eqnarray}
which are the expected results for a perfectly conducting wall.

For any finite value of $\epsilon$ it is not possible to simplify expression (\ref{Vepsf3}), as we did for the  $\epsilon\rightarrow\infty$ case (see equation (\ref{pcw})). However, one can obtain the $\epsilon$-dependence in a closed form for the asymptotic expressions for large ($x_0\gg 1$) and small ($x_0\ll 1$) distances. In these cases, it is possible to make the factorization: $V\left(z,\epsilon\right) \simeq g\left(\epsilon\right)V_0(z)$, where $g(\epsilon)$ is a function of $\epsilon$ that will be different for each distance limit. Directly from (\ref{Vepsf3}), we obtain
\begin{eqnarray}
\label{Veps_zS}
V\left(z,\epsilon\right) &\simeq& \frac{\hbar c}{32} \frac{\alpha_0 k_0}{z^3}\left[f^{\,\prime\prime}\left(1,\epsilon\right)-{1\over 2}\int_{-1}^1 dt f^{\,\prime\prime\prime}\left(\vert t\vert,\epsilon\right)\right]\;\;\;\;\;\;\;\;\;\;\;\;\;\;\;\;\;\;\;\;\;\;\;\;\;\;\;\;\;\;\;\;\;\; (x_0 \ll 1)
\\
%
\label{Veps_zL}
V\left(z,\epsilon\right) &\simeq& \frac{\hbar c}{32\pi} \frac{\alpha_0}{z^4}\left[2f\left(1,\epsilon\right)+f^{\,\prime}\left(1,\epsilon\right)+f^{\,\prime\prime}\left(1,\epsilon\right)-{1\over 2}\int_{-1}^1 {dt\over \vert t\vert} f^{\,\prime\prime\prime}\left(\vert t\vert,\epsilon\right)\right]\;\; (x_0 \gg 1)
\end{eqnarray}

In order to evaluate the above integrals, it is convenient to define $\kappa:=\epsilon - 1$ and then consider separately two cases, namely, $\kappa\ll 1$  and $\kappa$ very large. In the former, we expand $f^{\,\prime\prime\prime}(\vert t\vert,\kappa)$ in a power series of 
$\kappa$, while in the latter we expand $f^{\,\prime\prime\prime}(\vert t\vert,\kappa)$ in a power series of $1/\kappa$. These series are given, respectively, by:
\begin{eqnarray}
\label{f3kH}
f^{\,\prime\prime\prime}\left(\vert t\vert,\kappa\right)&=& {12\kappa\over t^4\vert t\vert}+\left(6-{30\over t^2}\right){\kappa^2\over t^4\vert t\vert}-\left(3+{15\over t^2}-{105\over 2t^4}\right){\kappa^3\over t^4\vert t\vert}+\mathcal O\left(\kappa^4\right)\;\;\;\;\; \left(\kappa \ll 1\right)
\nonumber \\
%
\label{f3kL}
f^{\,\prime\prime\prime}\left(\vert t\vert,\kappa\right)&=& {12\over t^4}{1\over \kappa^{1/2}} -{48\over t^4\vert t\vert}{1\over \kappa}+\left(18-{36\over t^4}+{120\over t^6}\right){1\over \kappa^{3/2}}+\mathcal O\left({1\over \kappa^2}\right)\;\;\;\;\;\;\;\;\; \left(\kappa \gg 1\right)
\nonumber
\end{eqnarray}
The above expressions are strongly divergent at $t=0$, which belongs to the interval of integration in both integrals appearing in equations (\ref{Veps_zS}) and (\ref{Veps_zL}). In order to circumvent this problem and extract a finite result from these ill defined integrals, we replace $t$ by $t\rightarrow t\pm i\delta$ and, after performing the integrations, we take $\delta\rightarrow 0$. Adopting this regularization prescription, we may write directly from equation (\ref{Veps_zS}) an expression for the interaction  between the atom and the semi-infinite dielectric medium for the case $x_0\ll 1$:
\begin{eqnarray}
\label{Vshort_eps}
V\left(z,\epsilon\right) &=& -\frac{\hbar \omega_0}{16} \frac{\alpha_0}{z^3}\kappa\sum_{n=0}^{\infty}\left(-1\right)^n\left({\kappa\over 2}\right)^n=-\frac{\hbar \omega_0}{8} \frac{\alpha_0}{z^3}\left({\epsilon - 1 \over \epsilon + 1}\right)
\end{eqnarray}

We should emphasize that the previous equation is valid for $\epsilon\ge 1$. In order to give a better understanding of this result, let us consider a sufficiently dilute medium so that the non-additivity of the van der Waals forces may be ignored. In this case, as a first approximation, one can use London's result \cite{London} for the interaction between two identical electrically polarizble atoms, $V_{AB}=-3\hbar\omega_0\alpha_0^2/4r^6$, to make a pairwise integration between the atom and the all other atoms that constitute the semi-infinite dielectric medium. Using the Clausius-Mosotti relation \cite{Jackson}, $4\pi N\alpha_0 = 3(\epsilon -1)/(\epsilon + 2)$, where $N$ is the number of atoms per unit volume of the medium, and assuming for a dilute medium that $\epsilon\approx 1$, one obtains  exactly half the value given by (\ref{Vshort_eps}) in first order in $\epsilon - 1$. This discrepancy is due to the fact that the London's result is valid only for distances smaller!
  than the wavelength of the dominant atomic transition and then can not be used for integration over 
all possible distances.

For the opposite limit, $x_0\gg 1$, the expression for $g(\epsilon)$ is not so simple like that written in (\ref{Vshort_eps}). However, it is not difficult to obtain series expansions in $1/\kappa$ or in $\kappa$. For large values of $\kappa$, we may write from (\ref{Veps_zL}):
\begin{equation}
\label{Vlarge_kappaL}
V\left(z,\kappa\right) = -\frac{3}{8\pi}\frac{\alpha_0\hbar c}{z^4}\left(1-{5\over 4\sqrt{\kappa}}+{22\over 15\kappa} - ...\right)
\end{equation}
which goes to result (\ref{V0large}) when $\epsilon \rightarrow \infty$. For small values of $\kappa$, it is possible to write:
\begin{equation}
\label{Vlarge_kappaS}
V\left(z,\kappa\right) = -\frac{23}{160\pi}\frac{\alpha_0\hbar c}{z^4}\kappa\left(1 - {169\kappa\over 322} + {2263\kappa^2 \over 7728}-...\right)
\end{equation}

Let us look at the previous equation more carefully. Following the same procedure as that adopted to obtain equation (\ref{Vshort_eps}), we can compute the interaction from the direct pairwise integration of the atom-atom potential in the Casimir-Polder limit: $V_{AB}\left(r\right)=-23\hbar c \alpha_0^2/4\pi r^7$. Setting $\epsilon \simeq 1$ and using again the Clausius-Mosotti relation, we may write for the integrated potential $V^{(int)}\left(z,\kappa\right)$:
\begin{equation}
\label{Vlarge_int}
V^{(int)}\left(z,\kappa\right) = -\frac{23}{160\pi}\frac{\alpha_0\hbar c}{z^4}\kappa\left(1 - {\kappa\over 3} + {\kappa^2 \over 9}-...\right)
\end{equation}

One can see from above that the leading terms of equations (\ref{Vlarge_kappaS}) and (\ref{Vlarge_int}) are the same. This means that in first order on $\kappa$ the interaction is due only to direct integration of the atom-atom potential in the Casimir-Polder limit. The higher order terms, however, are different because of the non-additivity of van der Walls forces so that the integration of a two-body potential can not give the exact result. A measure of this discrepancy may be given by the quantity:
\begin{equation}
\label{NonAdd}
{V\left(z,\kappa\right)-V^{(int)}\left(z,\kappa\right)\over V\left(z,\kappa\right)}= -{185 \over 966}\kappa+{303113\over 3732624}\kappa^2-{1325388223\over 39662862624}\kappa^3+...
\end{equation}
which essentially provides an average contribution of the many-body potentials to the interaction; this contribution is second order on $\kappa$. One may note also that differently of two-body potential that is always attractive, in this case the many-body contribution has a repulsive character (the many-body interaction may be repulsive depending on the geometry considered \cite{PowerThiru1985}).
\section{Conclusions}
In this work we studied the interacton of an atom with a semi-infinite dielectric medium characterized by a dielectric constant $\epsilon(\omega)$ which, in principle, can depend on the frequency. We used the general expressions for the energy level shifts of a small system coupled to a reservoir given by the master equation approach. For simplicity, we considered the atom (interpreted as the small system) as a two-level system and the radiation field (interpreted as the reservoir) in its vacuum state but 
submitted to the boundary conditions imposed by the presence of the semi-infinite dielectric medium. With these assumptions, we were able to establish a quite general expression for the interaction of the atom with the semi-infinite dielectric medium given by equations (\ref{Vepsk}-\ref{fepsk}). It is worth emphasizing that this result takes into account dispersion and is valid for all distance regimes.

We also analyzed the particular cases of a non-dispersive medium, where we obtained simple formulas for the interaction potential, specially in the retarded and non-retarded regimes. For these cases, it was possible to write the potential as a product of a function of $\epsilon$ only and the potential for a perfectly conducting medium in the corresponding limits. For the non-retarded limit, we obtained a simple analytical expression for the  atom-medium interaction valid for any $\epsilon$. For the retarded limit, however, simple expressions were obtained only for small values of $\kappa=\epsilon-1$ and large values of $\kappa$. Note that the first term in expansion (46), valid for large $\kappa$ and large distances coincides, as expected, with the Casimir and Polder result \cite{CasiPolder}. Further, for small $\kappa$ and retarded regime, we also computed the contribution of the many-body potentials to the interaction providing a measure of the non-additivity of the van de!
 r Waals interactions.
%
%
\footnotesize

{

\end{document}